\documentclass[final,5p,times,twocolumn]{elsarticle}

\usepackage{amssymb}
\usepackage{amsmath}
\usepackage{colortbl}
\definecolor{lightgrey}{rgb}{0.925, 0.925, 0.925}

\usepackage{mathabx}

\usepackage{graphicx}
\usepackage{latexsym}
\usepackage{times}
\usepackage[pagewise]{lineno}
\usepackage{hyperref}
\usepackage[most]{tcolorbox}
\usepackage{xcolor}
\usepackage{ctable}
\usepackage{pifont}

\usepackage{multirow}
\usepackage{dblfloatfix} 

\definecolor{darkgreen}{rgb}{0.0, 0.5, 0.0}
\usepackage{tikz,siunitx}
\usetikzlibrary{shapes.geometric,shapes.symbols}
\newcommand{\tikzsymbol}[2][circle]{\tikz[baseline=-0.5ex]\node[inner
sep=2pt,shape=#1,draw,#2]{};}%
\newcommand\eatpunct[1]{}
\usepackage{threeparttable}
\usepackage{lineno}

\journal{Energy and AI}

\begin{document}

\begin{frontmatter}



\title{Chain of Unit-Physics: A Primitive-Centric Approach to Scientific Code Synthesis}


\author[inst1]{Vansh Sharma} 

\author[inst1]{Venkat Raman}

%% Department of Aerospace Engineering,
\affiliation[inst1]{organization={University of Michigan},%Department and Organization
            %addressline={}, 
            city={Ann Arbor},
            postcode={48109-2102}, 
            state={MI},
            country={USA}}

%% Abstract
\begin{abstract}
\noindent\includegraphics[width=2.0\linewidth]{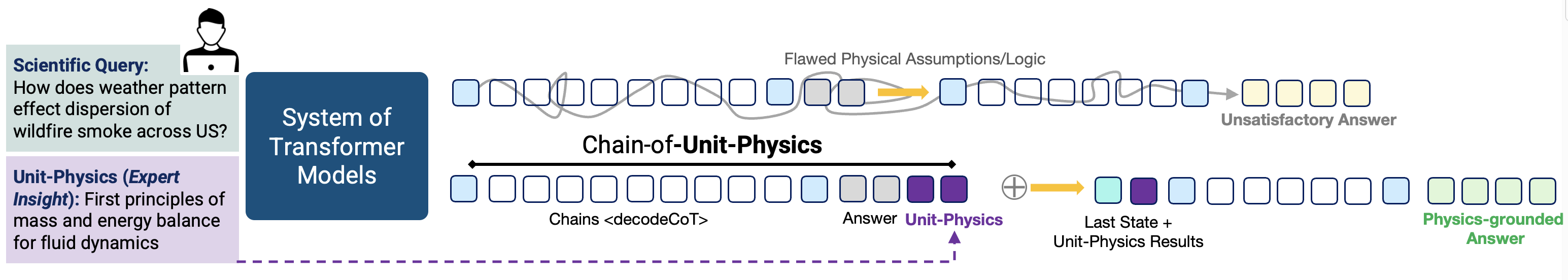}
Agentic large language models are proposed as autonomous code generators for scientific computing, yet their reliability in high-stakes problems remains unclear. Developing computational scientific software from natural-language queries remains challenging broadly due to (a) sparse representation of domain codes during training and (b) the limited feasibility of RLHF with a small expert community. 
To address these limitations, this work conceptualizes an inverse approach to code design, embodied in the Chain of Unit-Physics framework: a first-principles (or primitives)-centric, multi-agent system in which human expert knowledge is encoded as unit-physics tests that explicitly constrain code generation. The framework is evaluated on a nontrivial combustion task (12 degrees-of-freedom), used here as a representative benchmark for scientific problem with realistic physical constraints.
Closed-weight systems and code-focused agentic variants fail to produce correct end-to-end solvers, despite tool and web access, exhibiting four recurrent error classes: interface (syntax/API) hallucinations, overconfident assumptions, numerical/physical incoherence, and configuration fragility. Open-weight models with chain-of-thought (CoT) decoding reduce interface errors but still yield incorrect solutions. On the benchmark task, the proposed framework converges within 5–6 iterations, matches the human-expert implementation (mean error of 3.1$\times$10$^{-3}$\%), with a $\sim$33.4\% faster runtime and a $\sim$30\% efficient memory usage at a cost comparable to mid-sized commercial APIs, yielding a practical template for physics-grounded scientific code generation. As datasets and models evolve, zero-shot code accuracy will improve; however, the Chain of Unit-Physics framework goes further by embedding first-principles analysis that is foundational to scientific codes, thereby guiding more reliable and interpretable human–AI collaboration.
\end{abstract}

% \centering
% \includegraphics[width=.99\textwidth]{ graphical.png}

% \begin{highlights}
% \item Agentic LLMs remain systematically unreliable for high-stakes scientific workflows as code execution is often decoupled from scientific correctness.
% \item Failure modes of state-of-the-art agentic LLM systems evaluated on a nontrivial combustion benchmark task reveal four recurrent error classes in scientific code synthesis.
% \item Proposed framework encodes expert knowledge as unit-physics tests to constrain code generation - a fundamentally inverse-design approach.
% \item The framework reliably converges in 5–6 iterations, matches human-expert accuracy, and achieves lower memory usage at mid-sized commercial API–comparable cost.
% \end{highlights}

%% Keywords

\end{frontmatter}

% \linenumbers

\section{Introduction}
Emergence of large language models (LLMs) as key drivers of the integration of artificial intelligence (AI) across multiple domains, including computational science \cite{du2024largeLMCFD, yang2025large, sharma2024reliable}, have enabled intuitive human-like interactions with complex systems that outperform previous machine learning (ML) methods \cite{achiam2023gpt}. However, these models frequently exhibit unwarranted confidence, producing fluent but potentially erroneous outputs \cite{vanshRAG}, a liability that has serious implications in high-stakes engineering domains such as aircraft and engine design. In particular, during code generation scenarios \cite{openAI_codex,lu2024aiscientistfullyautomatedSakana}, LLMs might generate programs with correct syntactic structure \cite{yin2017syntactic} that may still violate logical requirements or drift from specified constraints over extended iterations \cite{sharma2025steering}. Such limitations underscore the urgent need for rigorous verification frameworks to ensure the accuracy and reliability of LLM-driven workflows.

From an engineering perspective, scientific code generation tasks diverge fundamentally from traditional software development. General purpose applications, such as mobile apps or utility libraries, prioritize rapid iteration, user experience, and flexibility, often tolerating minor glitches and relying on integration tests. In contrast, scientific software demands mathematically rigorous algorithms, bit‐for‐bit reproducibility, and exhaustive validation against analytical or reference benchmarks \cite{raman2019emerging}. Moreover, it must be optimized for large‐scale numerical workloads on high‐performance computing platforms without sacrificing stability \cite{sharma2025accelerating}, leveraging precision‐oriented languages and parallel libraries (e.g. MPI \cite{dongarra1995MPI}) to ensure both performance and accuracy. Given these divergent workflows and the pivotal role of testing in ensuring correctness, current research on LLM-driven code generation emphasizes on developing test cases from existing codebases~\cite{schafer2023empirical, paduraru2024llm_unittests_game, chen2024chatunitest} and transforming them into formal, verifiable unit-test specifications \cite{godage2025evaluating}. This also highlights that validation currently relies on comparing code outputs with predefined input-output datasets, an approach that is infeasible when problems are complex or due to limited number of experts in the field. In such domains, agents often generate erroneous codes that require extensive human debugging. In addition, for a given problem, there could be multiple correct code implementations, complicating validation based solely on output matching. While it may seem intuitive that providing unit tests alongside problem specifications would automatically improve the accuracy of LLM-generated code, this assumption has not been rigorously validated \cite{wang2024rocks}. The true effect of such test suites on the fidelity and robustness of model output remains largely unexplored \cite{mathews2024TDDLLM}, and, in particular, the mechanisms by which LLMs interpret, apply, and iteratively refine their code based on these tests~\cite{agarwal2025inputs} have yet to be systematically investigated.

\begin{figure}[t!]
\centering
\includegraphics[width=.48\textwidth]{ 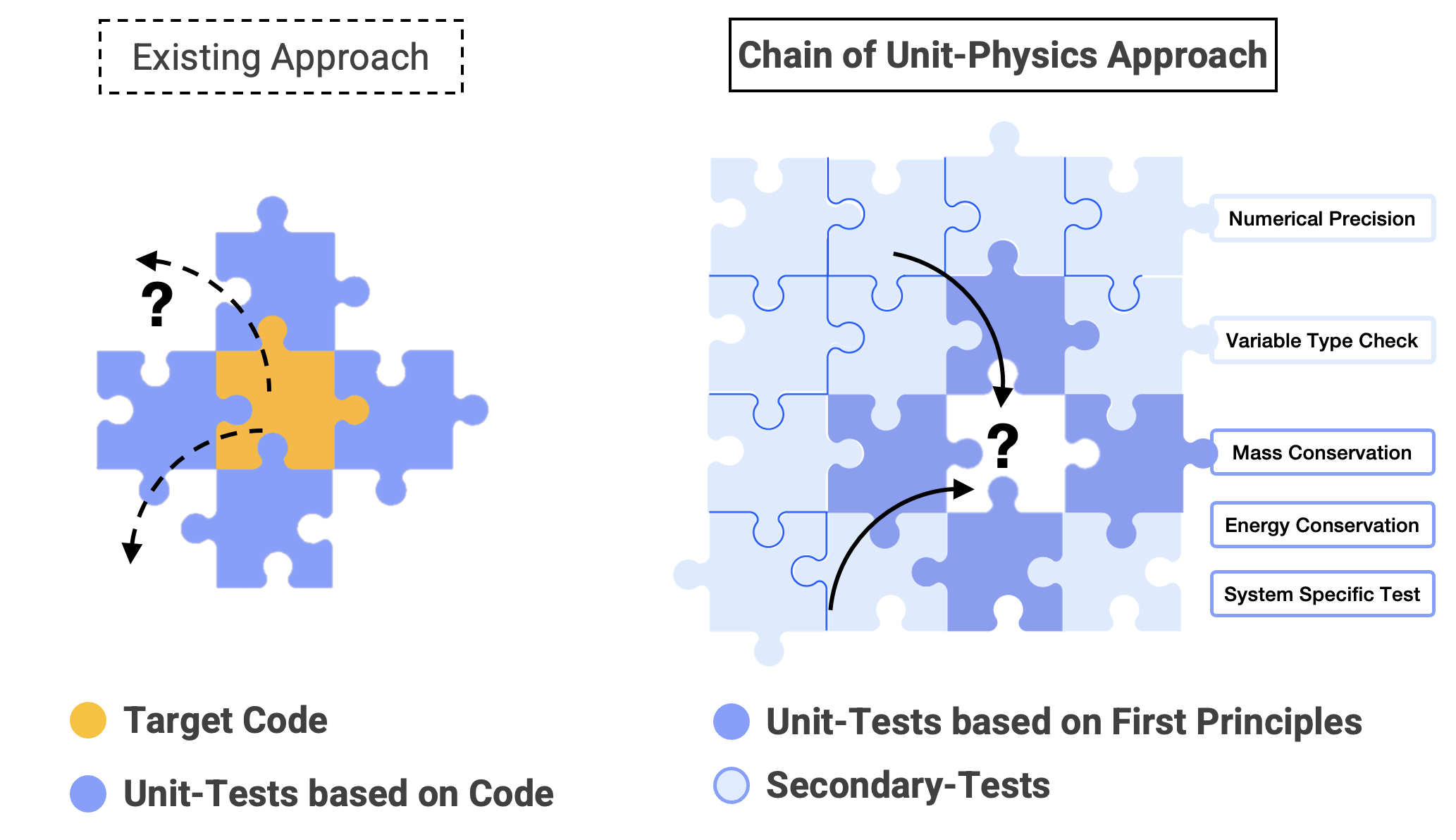}
\caption{Conceptual difference between Chain of Unit-Physics approach and existing methods. Left: the existing ``code-first" approach, where unit tests are written after implementation, merely exposing latent errors and forcing rework. Right: the proposed approach, in which a human expert specifies first-principles unit tests (e.g., conservation laws) that guide code generation.}
\label{fig:basic}
\end{figure}

Generating unit tests from existing code can inadvertently perpetuate latent errors~\cite{prasad2025learning}, particularly when those tests are not limited to generic checks, such as data type or format tests, rather emphasize algorithmic correctness~\cite{meta_unittestimprovement}. This issue is especially acute in the development of custom computational fluid dynamics (CFD) solvers \cite{sharma2024amrex, Zhang2026HighFidelity} built on libraries such as OpenFOAM \cite{jasak2009openfoam} and AMReX \cite{zhang2021amrex}. To overcome these limitations, this work systematically applies an inverse code design methodology (see Fig.~\ref{fig:basic}), formally known as test-driven development (TDD) \cite{beck2022TDD}, to scientific software in combustion science, one such domain where TDD approaches have not been thoroughly evaluated. This work proposes Chain of Unit-Physics, an approach that embeds human expert knowledge directly into distinct reasoning chains of the agentic system via ``unit-physics"—formalized, testable constraints that encode fundamental physics (e.g., energy conservation laws) and practitioner experience (e.g., dimensional / bound checks, and floating-point diagnostics).
This inverse-design method provides two key advantages in scientific software. First, human-authored tests embed deep domain expertise (first principles or primitives) into targeted validation checks, ensuring that each algorithmic component faithfully represents the underlying physics. Second, because these verification suites are authored by specialists, they impose stringent quality criteria - any failure of the LLM generated code then clearly indicates a gap in the expert's test specification itself, thereby shifting the source of error away from the model and onto the test suite or expert's own knowledge. 
This discussion naturally prompts the question:\emph{``If we can so precisely formalize our requirements, why not use a single numerical solver across all combustion applications, simply adapting for different fuel phases?"}. While a universal solver for all combustion problems in this context might seem appealing, it is neither practical nor conducive to the advancement of scientific discovery. There are a myriad of approaches to solving complex set of equations, including the choice of models, the choice of numerical schemes, their implementation as well as the specific regime of validity for all these choices.  Instead, designing targeted ``unit-physics" tests or ``primitives", each grounded in first principles, provides a more direct, transparent, and reliable framework for developing and verifying solver components for specific physical processes. Beyond the framework, a key contribution of this study is the systematic evaluation of human-authored unit tests to recast fundamental physics checks as stringent formal specifications for AI-driven code generation, thus reducing model ambiguity and minimizing error propagation. The following sections describe the proposed framework in detail ($\S$~\ref{sec:method}) and evaluate it on a benchmark scientific task ($\S$~\ref{sec:results}).

% Following the previous discussion, this study explores TDD in the context of AI-driven scientific code generation by incorporating unit tests developed by human experts. We propose a framework that integrates an LLM with a sandboxed execution environment, complete with user-specified libraries, to drive accurate code synthesis. The model generates multiple candidate solutions that ostensibly satisfy the problem statement, yet many initially fail the domain-specific ``unit physics" tests. An iterative feedback loop then captures the execution outputs and pinpoints failure modes, returning diagnostic information to the LLM for successive refinement. By continuously generating, testing, and correcting against formal physics specifications, this generate–test–refine cycle systematically identifies and fixes errors, progressively aligning the model’s outputs. By leveraging TDD paradigm with expert guided tests, the framework produces code that is functionally correct by design, quickly uncovers domain-specific flaws, and ensures reproducible and robust results, although contingent on the rigor of the test suite. Beyond the framework, a key contribution of this study is the systematic evaluation of human-authored unit tests to recast fundamental physics checks as stringent formal specifications for LLM-driven code generation, thus reducing model ambiguity and minimizing error propagation. 

\begin{figure*}[t!]
\centering
\includegraphics[width=.95\textwidth]{ 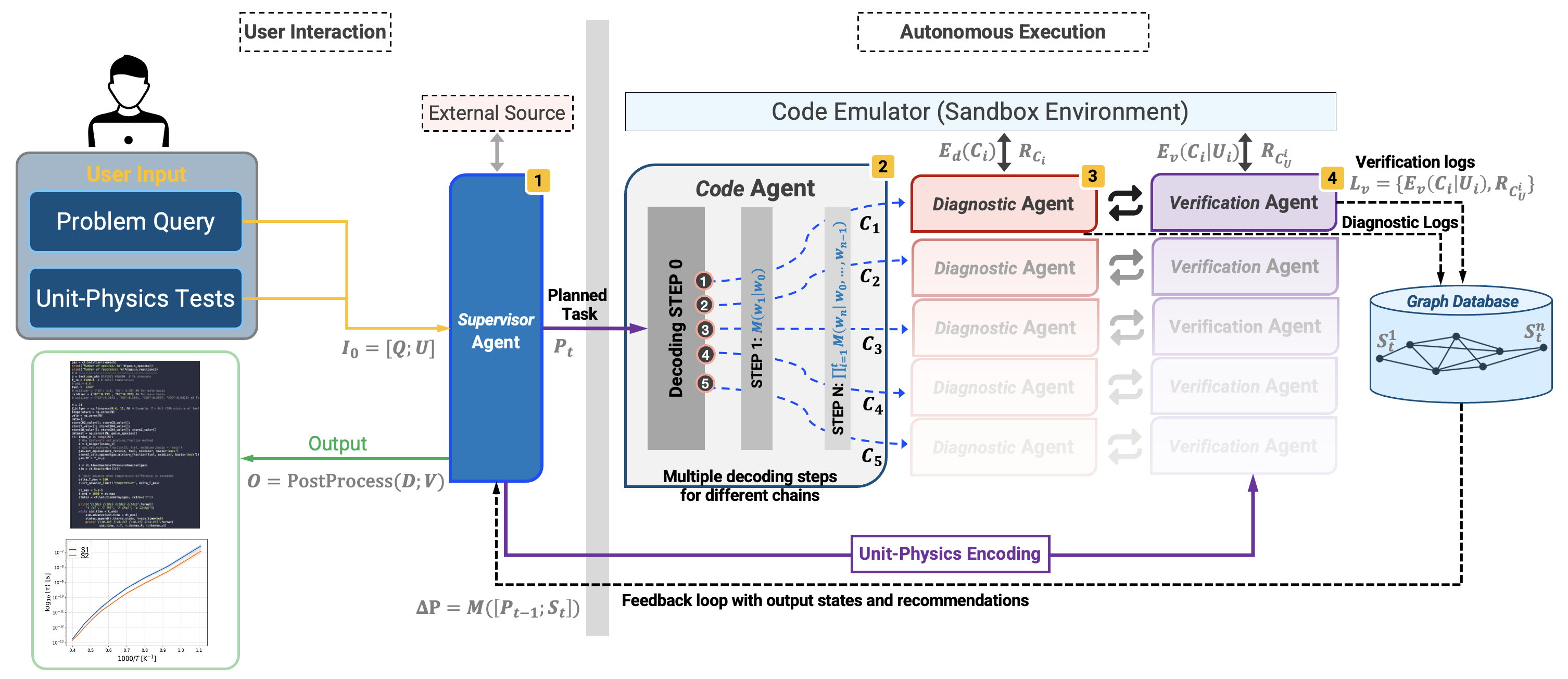}
\caption{Chain of Unit-Physics workflow: User queries and unit-physics tests are processed by a supervisor agent (1), which orchestrates chain-of-thought (CoT) code generation (2); diagnostic (3) and verification (4) agents then evaluate the code against physics-based tests and expert knowledge, feeding back signals that steer code synthesis toward physically consistent solutions. The final green‐bordered block confirms successful query execution.}
\label{fig:process_1}
\end{figure*}

\section{Methodology} \label{sec:method}
% \begin{figure}[t!]
% \centering
% \includegraphics[width=.48\textwidth]{ process_1.png}
% \caption{Workflow.}
% \label{fig:process_1}
% \end{figure}

\subsection{Multi-Agent System\label{sec:system}}

AI systems with higher degree of autonomy and goal-directed behavior, with a notion of planning and reasoning~\cite{kambhampati2024position}, embedded during model training, are termed as ``agents'' in this framework. While this definition is evolving, current work uses open-weight instruction-tuned decoder‑only LLMs as the base for agents: Llama 3.3 70B Instruct~\cite{grattafiori2024llama3herdmodels} and GPT-OSS-20B model with \texttt{o200k\_harmony} tokenizer~\cite{agarwal2025gpt}. Based on additional experiments (not shown), the framework adheres to established guidelines for the selection of model parameters (e.g. sampling temperature), specified in the respective reference articles~\cite{grattafiori2024llama3herdmodels, agarwal2025gpt}. The models are accessed using the \texttt{Transformers} library~\cite{wolf2020transformers_HF} and \texttt{vLLM} library~\cite{vLLM} while the agents are orchestrated through custom code within a Python 3.11 environment. Inference is performed with \texttt{PyTorch}~\cite{imambi2021pytorch} (GPU‑accelerated where available), while task decomposition and prompt management use custom code abstractions. The framework runs inside a dedicated Python virtual sandbox environment that is not pre‑configured with metadata on all available libraries. The agent is granted isolated code execution privileges within this sandbox, ensuring that any dependency installs or script executions cannot affect the system root directory or global files. During code execution, any missing dependencies are detected using builtin the \texttt{subprocess} library. Configuration and runtime states are managed with the \texttt{logging} library, and human-expert guidance can be integrated via command line interface (CLI). 

Figure~\ref{fig:process_1} illustrates a multi-agent LLM-driven scientific workflow that transforms a domain-specific user query into actionable code and visual results. In the input, the user's scientific question is paired with 'basis prompts' that establish execution permissions, tool availability, coding language settings, and transfer protocols. Additional input scopes the unit-physics tests that concatenated with the scientific query to form the complete input to the framework. The Supervisor Agent is the only part of the framework that interacts with the User. This agent identifies the objective by extracting critical parameters (e.g., thermodynamic variables, software/language requirements) and generates a structured plan that sequences the necessary steps and tool invocations and then sets up an autonomous iterative loop: each task is passed to Code Agent, which uses a custom decoding algorithm~\cite{wang2024chain} that elicits intermediate reasoning traces. The modified decoding algorithm branches only at the first generated token: instead of taking the greedy top-1 token, it spawns multiple paths by taking each of the top‑k candidates, then continues each path with standard greedy or beam search decoding. This procedure generates multiple code candidates per task similar to Chain-of-Thought (CoT)~\cite{wei2023CoTprompting}, without relying on explicit ``think step by step'' instructions. For every completed path, the algorithm identifies the answer span and computes a ``confidence" score as the average margin between the probabilities of the top-1 and the top-2 tokens over those answer tokens. These chains can be evaluated in parallel, either on separate GPUs or, depending on model memory requirements, jointly on a single GPU. While these self-assigned confidence scores enable chain selection via a user-specified confidence threshold, they are not a substitute for validation against first-principles physical constraints and numerical consistency checks.

The pruned candidate chains are then asynchronously routed to two downstream agents, a Diagnostic agent and a Verification agent, which together constitute ``chain of unit-physics". The Diagnostic Agent first performs preliminary checks (e.g. dependency installation) by automatically running code in the emulator, analyzing any errors, and applying targeted corrections. Subsequently, the Verification Agent applies formalized unit-physics tests to assess physical and numerical consistency. These primitives yield physics-grounded verification even without reference datasets. At each stage of the framework, the agent states and logs for different interactions persist in a graph-based database. The database stores the summarized execution and diagnostic logs for each code candidate on nodes, to avoid context-window limitations~\cite{vanshRAG}, while the edges define the logic route and transitions. If any candidate code cannot be corrected within a user-specified number of attempts, the complete error log and the candidate implementation are returned to the Supervisor Agent for updating the plan as needed, optionally incorporating human feedback. Finally, the agent consolidates the output into domain-relevant visualizations (for example, line graphs and contour graphs of key quantities of interest), closing the loop between natural-language inquiry and quantitative scientific insight. 

The proposed framework is formally described as follows: let $M$ be the decoder‑only Transformer model with 
$Q = \mathrm{Tokenize}(\text{UserQuery})$ and 
$U = \mathrm{Tokenize}(\text{UnitPhysicsTests})$ 
concatenated into the initial supervisor input 
$I_{0} = [\,Q;\,U\,]$.  
The sandboxed execution operator $\mathrm{Exec}(\cdot)$ returns a pair $(R,E)$ where
$R$ is the runtime output (or $\varnothing$ on failure or no valid output) and
$E$ is the error message (or $\varnothing$ if successful or no error message).  
Then the workflow is:

\begin{itemize}
  \item \textbf{Supervisor Initialization.}
    \begin{equation}
      \label{eq:super-init}
      P_{0} \;=\; M\bigl(I_{0}\bigr)
    \end{equation}

  \item \textbf{Multi‑Chain Code Decoding (Code Agent).}
    For iterations $t = 1,2,\dots$ the supervisor produces a decoding prompt $P_{t-1}$ and the
    code agent generates $K$ candidate programs via multi‑chain decoding:
    \begin{equation}
      \label{eq:multichain}
      \{C_{t}^{(k)}\}_{k=1}^{K}
      \;=\;
      \mathrm{CoTDecode}_{K}\bigl(M, P_{t-1}\bigr)
    \end{equation}
    where each chain $k$ branches at the first token.

  \item \textbf{Code Execution.}
    For each candidate chain $k$:
    \begin{equation}
      \label{eq:exec-sandbox}
      \bigl(R_{t}^{(k)},\,E_{t}^{(k)}\bigr)
      \;=\;
      \mathrm{Exec}\bigl(C_{t}^{(k)}\bigr)
    \end{equation}

  \item \textbf{Diagnostics (Diagnostic Agent).}
    \begin{equation}
      \label{eq:diag}
      D_{t}^{(k)}
      \;=\;
      \mathrm{Diag}\bigl(C_{t}^{(k)}, R_{t}^{(k)}, E_{t}^{(k)}\bigr)
    \end{equation}

  \item \textbf{Testing Against Unit‑Physics (Verification Agent).}
    \begin{equation}
      \label{eq:verify}
      \bigl(V_{t}^{(k)},\,\hat{E}_{t}^{(k)}\bigr)
      \;=\;
      \mathrm{Verify}\bigl(C_{t}^{(k)}, U\bigr)
    \end{equation}

  \item \textbf{Logging \& State Aggregation (Database).}
    \begin{equation}
      \label{eq:logs}
      L_{d}^{(t)} = \{(C_{t}^{(k)}, R_{t}^{(k)}, D_{t}^{(k)})\}_{k=1}^{K},
      \qquad
      L_{v}^{(t)} = \{(C_{t}^{(k)}, V_{t}^{(k)}, \hat{E}_{t}^{(k)})\}_{k=1}^{K}
    \end{equation}
    \begin{equation}
      \label{eq:state}
      S_{t} \;=\; \mathrm{Summarize}\bigl(L_{d}^{(t)}, L_{v}^{(t)}\bigr)
    \end{equation}

  \item \textbf{Supervisor Plan Update.}
    With optional external guidance $H_{t}$:
    \begin{equation}
      \label{eq:plan-update}
      \Delta P_{t} = M\bigl([P_{t-1};\,S_{t};\,H_{t}]\bigr),
      \quad
      P_{t} = \mathrm{Refine}\bigl(P_{t-1},\,\Delta P_{t}\bigr)
    \end{equation}

  \item \textbf{Termination \& Final Output.}
    If some chain $k^{\star}$ satisfies diagnostic and verification criteria:
    \begin{equation}
      \label{eq:select}
      k^{\star}
      \;=\;
      \arg\max_{k} \mathrm{Score}\bigl(D_{t}^{(k)}, V_{t}^{(k)}\bigr),
      \quad
      O = \mathrm{PostProcess}\bigl(R_{t}^{(k^{\star})}\bigr)
    \end{equation}
    otherwise, continue with iteration $t+1$.
\end{itemize}

\noindent\textbf{Notation:}
\begin{itemize}
  \item $\mathrm{CoTDecode}_{K}(M,P)$: multi‑chain decoding that branches on the first token
        and greedily completes $K$ candidate programs under prompt $P$.
  \item $\mathrm{Diag}$: diagnostic analysis of runtime behavior and errors.
  \item $\mathrm{Verify}$: execution of candidate code against unit‑physics tests $U$.
  \item $\mathrm{Summarize}$: compression of execution and verification logs into a state $S_{t}$.
  \item $\mathrm{Refine}(P,\Delta P)$: update of the supervisor plan.
  \item $\mathrm{Score}$: ranking function combining diagnostic and verification signals.
  \item $\mathrm{PostProcess}(R)$: formatting, data manipulation, and visualization of the selected result.
\end{itemize}

\subsection{Unit-Physics Encoding\label{sec:prim}}
Unit-physics primitives encode the first-principles constraints that a combustion
expert would impose on any admissible thermochemical state. These constraints
may be problem-specific (e.g., detonations vs.\ constant-pressure reactors) or
generic to reacting mixtures, and are designed to be portable across mechanisms
and numerical implementations (\textit{primitives are domain‑portable}). In
contrast to conventional unit tests, which mostly compare implementation details
against reference outputs, primitives enforce conservation, thermodynamic
consistency, and physical consistency, and thus provide partial verification
even when no ground‑truth solution is available.

For example, in a problem consisting of a spatially homogeneous (zero‑dimensional) reactor that integrates species mass fractions $Y_i$, temperature $T$, pressure $p$, and density $\rho$, typical primitives include:
\begin{itemize}
    \item \textbf{Species mass conservation:}
      {\footnotesize $\bigl|\sum_i Y_i - 1\bigr| \leq \epsilon$, with
      $\epsilon = 10^{-16}$}.
    \item \textbf{Equation‑of‑state residual for an ideal mixture:}
      {\footnotesize $\bigl|\, p - \rho R T / \bar{W} \,\bigr| \leq \epsilon$},
      where $\bar{W}$ is the molecular weight of the mixture.
    \item \textbf{Physical bounds:}
      {\footnotesize $1 \geq Y_i \geq 0$, $T_{\min} \leq T \leq T_{\max}$,
      $p>0$, $\rho>0$}.
    \item \textbf{Conservation of inert diluents (e.g.\ nitrogen):}
      {\footnotesize $\bigl|Y_{\mathrm{N_2}}^{(1)} - Y_{\mathrm{N_2}}^{(2)}\bigr|
      \leq \epsilon$} between two states of the same closed system.
    \item \textbf{Dimensional consistency of thermochemical quantities:}
      {\footnotesize $h_i$ in J\,kg$^{-1}$, $\hat{\omega}_i$ in
      kg\,m$^{-3}$\,s$^{-1}$} for species enthalpy and mass production rate.
\end{itemize}

Similarly, for a one‑dimensional Zeldovich–von Neumann–Döring (ZND) detonation
structure~\cite{zel1940theory,vonNeumann_ADB967734,Doering_Burkhard_1949},
additional process‑specific primitives supplement these generic checks.
Denoting pre‑ and post‑shock states by superscripts $^{(1)}$ and $^{(2)}$ and
$v = 1/\rho$ as specific volume, we impose:
\begin{itemize}
    \item \textbf{Rankine–Hugoniot energy closure:}
      {\footnotesize
      $\left|\, h_2 - h_1 + \tfrac{1}{2}(p_2+p_1)(v_2-v_1) \,\right| \leq
      \epsilon$}.
    \item \textbf{Chapman–Jouguet (CJ) condition at the equilibrium product
      state:} {\footnotesize $\mathrm{Mach}_{\mathrm{CJ}} = 1$}.
    \item \textbf{Consistency of product states between ZND and equilibrium
      (HP) calculations:}
      {\footnotesize
      $\left|\dfrac{T^{\mathrm{ZND}} - T^{\mathrm{HP}}}{T^{\mathrm{HP}}}\right|
      \leq \epsilon$} and
      {\footnotesize $\bigl|\chi_i^{\mathrm{ZND}} - \chi_i^{\mathrm{HP}}\bigr|
      \leq \epsilon$} for major species mass or mole fractions $\chi_i$.
\end{itemize}

These primitives can be enforced during both synthesis and evaluation of
agent‑generated codes, providing thermochemically grounded verification signals
even in the absence of curated reference datasets.  Within the AI framework,
they may be supplied either as natural‑language descriptions or as a structured
JSON specification that is parsed into executable checks.

\section{Results} \label{sec:results}
This study will consider a canonical reactor-design problem (prompt shown below) to calculate the ignition delay time (IDT) for two different types of systems: 1. Closed-weight models and 2. Open-weight models. Although the Cantera library~\cite{cantera} provides built-in routines to perform this calculation, the system is deliberately prompted to implement the time integrators directly for the full $N+1$ degree-of-freedom (dof) problem (12 dofs for H$_2$), rather than calling these high-level functions. The fundamental nature of this task makes it well suited to illustrate why unit-physics must be explicitly encoded in the system, and the same lesson extends to analogous problems in other scientific domains. Since this is a domain-specific realistic problem with nontrivial numerical constraints and the available benchmarks do not resolve such fine‑grained combustion physics or numerical aspects, there is limited information on the model behavior here. The present case therefore serves as a proxy for real engineering problems that fall outside standard benchmark coverage. The closed-weight models are evaluated using their APIs, and open-weight models are distributed on 4 NVIDIA H100 GPUs based on workload.

\begin{tcolorbox}[title = Prompt for Zero-D Reactor Task, colback=blue!5!white,colframe=blue!55!black, fonttitle=\bfseries, width = 0.99\linewidth,]
\phantomsection
\label{box:reactor}
\footnotesize{You need to develop an ignition delay calculator for different fuels.
They key here is to provide an explicit Euler or RK scheme to integrate states and using Cantera lib in python.
STRICT: You cannot use Cantera's reactor functions. You can only obtain gas object related properties from Cantera. 
For starters you can use fuel as hydrogen at temperature of 1300 Kelvin at pressure of 101325 Pa for stoichiometric composition. }
% \label{prompt_reactor}
\end{tcolorbox}

% \cellcolor{green!30}\ding{51}
% \cellcolor{orange!30}\ding{119}
% \cellcolor{red!30}\ding{55}
\begin{table*}[!tbp]
    \centering
    \caption{Results for different closed-weight models and agents. The possible outcomes are: \ding{51} indicates successful output, \ding{119} represents partially successful output, and \ding{55} signifies a failure.}
    \resizebox{2.1\columnwidth}{!}{\begin{tabular}{|c|l|c|c|}
        \specialrule{.2em}{.1em}{.1em}
        \textbf{Model} & \textbf{Output} & \textbf{\textbf{Failed Step}} & \textbf{Evaluation} \\
        \specialrule{.2em}{.1em}{.1em}
        
        ChatGPT & CanteraError: gas object has no attribute \texttt{`int\_energies\_mass'} & Derivative & \cellcolor{red!30}\ding{55} \\
        Sonnet 4.5 & CanteraError: Phase::setTemperature T = -56199.3659 & RK integration & \cellcolor{red!30}\ding{55} \\
        Gemini 2.5 Pro & CanteraError: \texttt{h2\_gri30.yaml} not found & Incorrect inputs & \cellcolor{red!30}\ding{55} \\
        \specialrule{.2em}{.1em}{.1em}
        
        \textbf{AI System} & \textbf{Output} & \textbf{\textbf{Failed Step}} & \textbf{Evaluation} \\
        \specialrule{.2em}{.1em}{.1em}
         Codex v0.44.0 & CanteraError: gas object has no attribute \texttt{`enthalpies\_mass'} & Derivative & \cellcolor{red!30}\ding{55} \\
        Claude Code v2.0.5 & CanteraError: Phase::setTemperature T is negative & RK integration & \cellcolor{red!30}\ding{55} \\
        \specialrule{.2em}{.1em}{.1em}
    \end{tabular}}
    
    \label{tab:test_close}
\end{table*}

\subsection{Closed-weight Models and Systems\label{sec:closed}}

First, closed-weight models such as ChatGPT\cite{openai2025chatgpt}, Claude Sonnet 4.5~\cite{anthropic2025claude} and Gemini 2.5 Pro~\cite{google2025gemini25} are tested for the given reactor prompt. All models have CoT, have access to local Python emulator to verify the codes\footnote{Codes to be released after paper is accepted.} and can access the Web during script generation. 
The results in the upper half of Table~\ref{tab:test_close} show that none of the evaluated models produced the target code, and detailed analysis highlights an incorrect understanding of ignition delay time, however, which does not impede code execution.  

\textbf{\textit{OpenAI ChatGPT}:} The function \texttt{derivatives()} that computes derivatives of state variables with time fails on execution. The code shown in the snippet exemplifies an API hallucination: it calls a nonexistent gas-object method (reported as \texttt{int\_enegries\_mass}), which is a clear syntax/API error based on Cantera documentation. An additional fundamental error lies in the process assumption: the model treats the evolution as a constant-pressure process rather than the intended constant-volume process. Furthermore, the model confidently suggests using \texttt{gri30.yaml} mechanism (CH$_4$ focused) for H$_2$ combustion as it will contain relevant reactions. 
\begin{tcolorbox}[title = \footnotesize Code snippet from ChatGPT: \texttt{derivatives()}, colback=black!5!white,colframe=gray!75!black, fonttitle=\bfseries, width = 0.99\linewidth,]
\phantomsection
\label{box:openai}
\footnotesize{ $\dots$ \textcolor{gray}{\# mass fractions} \\
    Y = s\_massdens / rho

    \textcolor{gray}{\# update gas state so Cantera returns consistent properties} \\
    gas.TPY = T, P, Y  \# T [K], P [Pa], mass fractions Y

    \textcolor{gray}{\# molar net production rates [kmol/m3/s], shape (n\_species,)} \\
    omega\_dot = gas.net\_production\_rates  \# kmol / m3 / s

    \textcolor{gray}{\# molecular weights [kg/kmol]} \\
    W = gas.molecular\_weights  \# kg / kmol

    \textcolor{gray}{\# mass production rates [kg/m3/s]} \\
    mdot = W * omega\_dot  \# elementwise, kg/m3/s

    \textcolor{gray}{\# specific internal energies of species [J/kg]} \\
    u\_species = gas.int\_energies\_mass  $\leftarrow$ \textbf{\textcolor{red}{Error}}

    \textcolor{gray}{\# mixture constant-volume specific heat [J/kg/K]} \\
    cv\_mix = gas.cv\_mass

    \textcolor{gray}{\# dT/dt from internal energy balance (closed homogeneous system}) \\
    numerator = np.dot(u\_species, mdot)  \# J/m3/s \\
    dTdt = - numerator / (rho * cv\_mix) $\dots$ $\leftarrow$ \textbf{\textcolor{red}{Error}}}
% \label{prompt_reactor}
\end{tcolorbox}

\textbf{\textit{Anthropic Claude Sonnet}:} The model yields the most elaborate deliverables: an algorithmic data-flow outline, a report, and an instructions file, even though such artifacts were not requested in the prompt. This likely reflects internal prompt expansion or meta-scaffolding~\cite{carpineto2012survey}, where a smaller model expands the query with detailed problem specifications to reduce hallucinations. The model selects the correct reaction mechanism (\texttt{h2o2.yaml}), but the final code implementation was incorrect: during RK time integration, it produces a negative temperature, indicating numerical/physical instability. After an expert review of the code and additional tests with smaller time steps, a second issue (similar to ChatGPT's code) was identified: the gas state was being set using an incompatible combination of thermodynamic formulation. The code snippet shows a function that is formulate by the model to update states for a constant volume system, but it currently applies a constant-pressure energy formulation which is highly incorrect. Instead of using species enthalpies ($h_k$) and $c_p$, it should use species internal energies and mixture $c_v$. Thus, even if the RK step had succeeded, the incorrect state-setting logic would have caused a subsequent failure. 
\begin{tcolorbox}[title = \footnotesize Code snippet from Sonnet: \texttt{get\_derivatives\_const\_vol()}, colback=black!5!white,colframe=gray!75!black, fonttitle=\bfseries, width = 0.99\linewidth,]
\phantomsection
\label{box:sonnet}
\footnotesize{ $\dots$\textcolor{gray}{\# set gas state: temperature, density, mass fractions} \\
gas.TDY = T, rho, Y  \quad \textcolor{gray}{\# T [K], $\rho$ [kg/m$^3$], Y mass fractions}

\textcolor{gray}{\# molar net production rates [kmol/m$^3$/s]} \\
omega = gas.net\_production\_rates

\textcolor{gray}{\# molecular weights [kg/kmol]} \\
W = gas.molecular\_weights

\textcolor{gray}{\# mass-fraction time derivatives [1/s]} \\
dYdt = omega * W / rho  \quad \textcolor{gray}{\# elementwise multiply}

\textcolor{gray}{\# species specific enthalpies [J/kg]} \\
h = gas.partial\_molar\_enthalpies / W $\leftarrow$ \textbf{\textcolor{red}{Error}}

\textcolor{gray}{\# mixture constant-pressure specific heat [J/kg/K]} \\
cp = gas.cp\_mass

\textcolor{gray}{\# dT/dt from enthalpy balance (closed homogeneous system)} \\
dTdt = - \text{np.dot}(h, omega * W) / (rho * cp) $\dots$ $\leftarrow$ \textbf{\textcolor{red}{Error}} }

% \label{prompt_reactor}
\end{tcolorbox}

\textbf{\textit{Google Gemini}:} In comparison to other models, Gemini did capture the correct process (constant-volume) assumption but fails at input initialization, requesting a mechanism file, \texttt{h2\_gri30.yaml}, that does not exist. Upon manually fixing this error, the code still shows erroneous outputs: shown in the snippet is a condition that masks a problem - if Cantera raises a state error, the code will allow the solver to continue and silently provide misleading output after that point. Non-experts using these models to write scientific code need extreme caution: models may make fundamentally wrong physical assumptions and mask them, even if they follow the provided documentation.
\begin{tcolorbox}[title = \footnotesize Code snippet from Gemini: \texttt{calculate\_derivatives()}, colback=black!5!white,colframe=gray!75!black, fonttitle=\bfseries, width = 0.99\linewidth,]
\phantomsection
\label{box:gemini}
\footnotesize{ $\dots$ try:\\
    gas.TPY = T, P, Y \\
except Exception as e:\\
    \textcolor{gray}{\# Handle cases where the state is invalid (e.g., T < 0)}\\
    print(f"Error setting Cantera state: \{e\}") \\
    \textcolor{gray}{\# Return zeros to avoid crashing the solver}\\
    return np.zeros\_like(y\_state) $\dots$ $\leftarrow$ \textbf{\textcolor{red}{Error}} }

% \label{prompt_reactor}
\end{tcolorbox}

Since the focus is on AI systems, we also evaluate agentic deployments of these models (see Table~\ref{tab:test_close}). The agentic variants: Codex~\cite{openAI_codex} and Claude Code~\cite{anthropic2025claudeCODE} exhibit the same failure patterns observed in their base models - given that the agent orchestrations are layered on the same underlying models despite having access to Web. The Claude Code agent ignores the instruction to implement an Euler integrator, instead providing only an RK4 scheme with faulty adaptive time-stepping logic and the same thermodynamic inconsistency observed for Sonnet 4.5. Additionally, the Codex agent repeats the same syntactic mistake as the GPT model. The agents systems are able to navigate and work with files in local directories, creating requested files in the correct locations, but still end up generating erroneous code. Essentially, workflow or agentic system does not remediate core model/API errors as observed previously. Furthermore, across models and systems, we observe a systematic mechanism mis-specification: the code tends to default to \texttt{gri30.yaml} even for H$_2$ fuel, which undermines fidelity from the outset. Despite being trained by different companies, the models select mechanisms based on what is most documented and frequently co-occurs with terms such as ``Cantera" and ``ignition," not on physical fidelity, so they all tend to converge on the same overrepresented mechanism in their training data distribution.

Collectively, these outcomes illustrate four recurring error classes in agentic code synthesis for scientific workflows: (i) interface hallucinations (nonexistent methods/attributes), (ii) over-assumption about scientific process (hard assumptions that do not translate to correct codes), (iii) numerical and physical incoherence (instabilities such as negative T and misuse of thermodynamic state variables), and (iv) configuration fragility (missing files and unsuitable default mechanisms). 
%These failures explain why no model achieved the target implementation and highlight the need for physics- and API-grounded validation ``primitives'' prior to execution.

% \cellcolor{green!30}\ding{51}
% \cellcolor{orange!30}\ding{119}
% \cellcolor{red!30}\ding{55}
\begin{table*}[!hbt]
    \centering
    \caption{Results for different open-weight models and agent. See Table~\ref{tab:test_close} for outcome description.}
    \resizebox{2.1\columnwidth}{!}{\begin{tabular}{|c|l|c|c|}
        \specialrule{.2em}{.1em}{.1em}
        \textbf{Model} & \textbf{Output} & \textbf{\textbf{Failed Step}} & \textbf{Evaluation} \\
        \specialrule{.2em}{.1em}{.1em}
        
        Llama3.3-70B & Error in RK4 function: Not-a-Number (\texttt{NaN}) outputs & RK integration & \cellcolor{red!30}\ding{55} \\
        Llama3.3-70B + CoT & Incorrect definition of IDT (time where \texttt{gas.T} $\geq$ 100 + T$_0$) & IDT logic &\cellcolor{orange!30}\ding{119} \\
        OSS-20B & CanterError: syntax \texttt{`set\_mass\_fractions'} and \texttt{gas.T} is not writeable & Inputs \& RK & \cellcolor{red!30}\ding{55} \\
        OSS-20B + CoT & Error: dTdt wrong size. Occurs as \texttt{gas.enthalpy\_mass} is assumed for each specie & Incorrect logic & \cellcolor{red!30}\ding{55} \\
        \specialrule{.2em}{.1em}{.1em}

        \textbf{AI System} & \textbf{Output} & \textbf{\textbf{Failed Step}} & \textbf{Evaluation} \\
        \specialrule{.2em}{.1em}{.1em}
         CoT+Unit Phy (AI+Human) & Ignition delay time: 1.13179e-05 s & - &  \cellcolor{green!30}\ding{51} \\
        \specialrule{.2em}{.1em}{.1em}
        
    \end{tabular}}
    
    \label{tab:test_open}
\end{table*}
\subsection{Open-weight Models and Systems}

Open-weight models include the Llama (sampling temperature of 0.6) and OSS models (sampling temperature of 0.3 and reasoning effort set to medium), and are accessed using the framework described in $\S$\ref{sec:method}. Two types of tests are performed here: first with the standard model and second with model coupled to CoT decoding algorithm to improve program search efficiency. During both these tests, the models do not have access to the Web or a python emulator. From top row of Table~\ref{tab:test_open}, across all models, the synthesized codes\footnote{Codes to be released after paper is accepted.} fail on basic physics/API checks. Llama-3.3-70B produces \texttt{NaN} states during the RK4 step, indicating  numerical/physical instability during the integration process. Adding CoT does not fully resolve such errors; while output code executes, but the model instead mis-defines IDT (defined by the model as the time when T$\geq$T+100), a conceptual or semantic error that runs the code but produces the wrong output. OSS-20B without CoT already fails at input handling and state initialization: it invokes an invalid Cantera call and overwrites the gas state in two incorrect ways: (i) treating mole fractions as mass fractions and (ii) calling the wrong function for setting for mass fractions. With CoT enabled, the failure mode shifts to inconsistent thermodynamics: the model implicitly treats the scalar \texttt{gas.enthalpy\_mass} as a species-resolved vector, producing a temperature derivative of the wrong shape (see code snippet). %Overall, the dominant failure steps show the model's reliance on internal knowledge gained through pre-training is not sufficient and often leads to syntactical errors/API misuses. 
Overall, CoT shifts the dominant failure mode from calling nonexistent functions to misusing valid APIs—an improvement, but still insufficient. Hence, one CoT-based model is marked `partially successful': the codes executed without any API errors, but the IDT compute logic was incorrect. We also note the same mechanism mis-specification (use of \texttt{gri30.yaml} for H$_2$ fuel) in these tests as well.

\begin{tcolorbox}[title = \footnotesize Code snippet from OSS-20B (no CoT): \texttt{init()}, colback=black!5!white,colframe=white!55!black, fonttitle=\bfseries, width = 0.99\linewidth,]
\phantomsection
\label{box:oss_noCoT}
\footnotesize{ $\dots$
\textcolor{gray}{\# create gas with GRI-Mech 3.0 (H$_2$/O$_2$/N$_2$ chemistry)} \\
gas = ct.Solution('gri30.yaml')

\textcolor{gray}{\# set initial thermodynamic state: T [K], P [Pa]} \\
gas.TP = 1300.0, 101325.0  \quad \textcolor{gray}{\# 1300 K, 1 atm}

\textcolor{gray}{\# stoichiometric H$_2$/air mixture: H$_2$ + 0.5 O$_2$ + 3.76 N$_2$} \\
\textcolor{gray}{\# unnormalized “masses” proportional to m\_{H2} = 1.0, m\_{O2} = 0.5, m\_{N2} = 3.76} \\
masses = \text{np.array}([1.0, 0.5, 3.76]) \\[4pt]
\textcolor{gray}{\# normalize to sum to 1 (convert to mass fractions)} \\
masses /= masses.\text{sum}() \\[4pt]
\textcolor{gray}{\# set mass fractions in Cantera} \\
gas.set\_mass\_fractions(masses)
$\dots$ $\leftarrow$ \textbf{\textcolor{red}{Error}} }

% \label{prompt_reactor}
\end{tcolorbox}

\begin{tcolorbox}[title = \footnotesize Code snippet from OSS-20B (CoT): \texttt{rhs()}, colback=black!5!white,colframe=white!55!black, fonttitle=\bfseries, width = 0.99\linewidth,]
\phantomsection
\label{box:oss}
\footnotesize{ $\dots$
T = y[0] \\
X = y[1:] \\[4pt]
gas.TPX = T, gas.P, X \\[4pt]
rho = gas.density \\ 
cv  = gas.cp\_mass - ct.gas\_constant / gas.mean\_molecular\_weight \\[4pt]
h = gas.enthalpy\_mass $\leftarrow$ \textbf{\textcolor{red}{Error}} \\[4pt] 
w = gas.net\_production\_rates \\
W = gas.molecular\_weights \\[4pt]
dTdt = - (1.0 / (rho \* cv)) \* \text{np.dot}(h, w) \\
dXdt = (1.0 / rho) \* (w / W)
$\dots$  }
% \label{prompt_reactor}
\end{tcolorbox}

\subsubsection{Chain of Unit-Physics System}

\begin{figure}[t!]
\centering
\includegraphics[width=.35\textwidth]{ 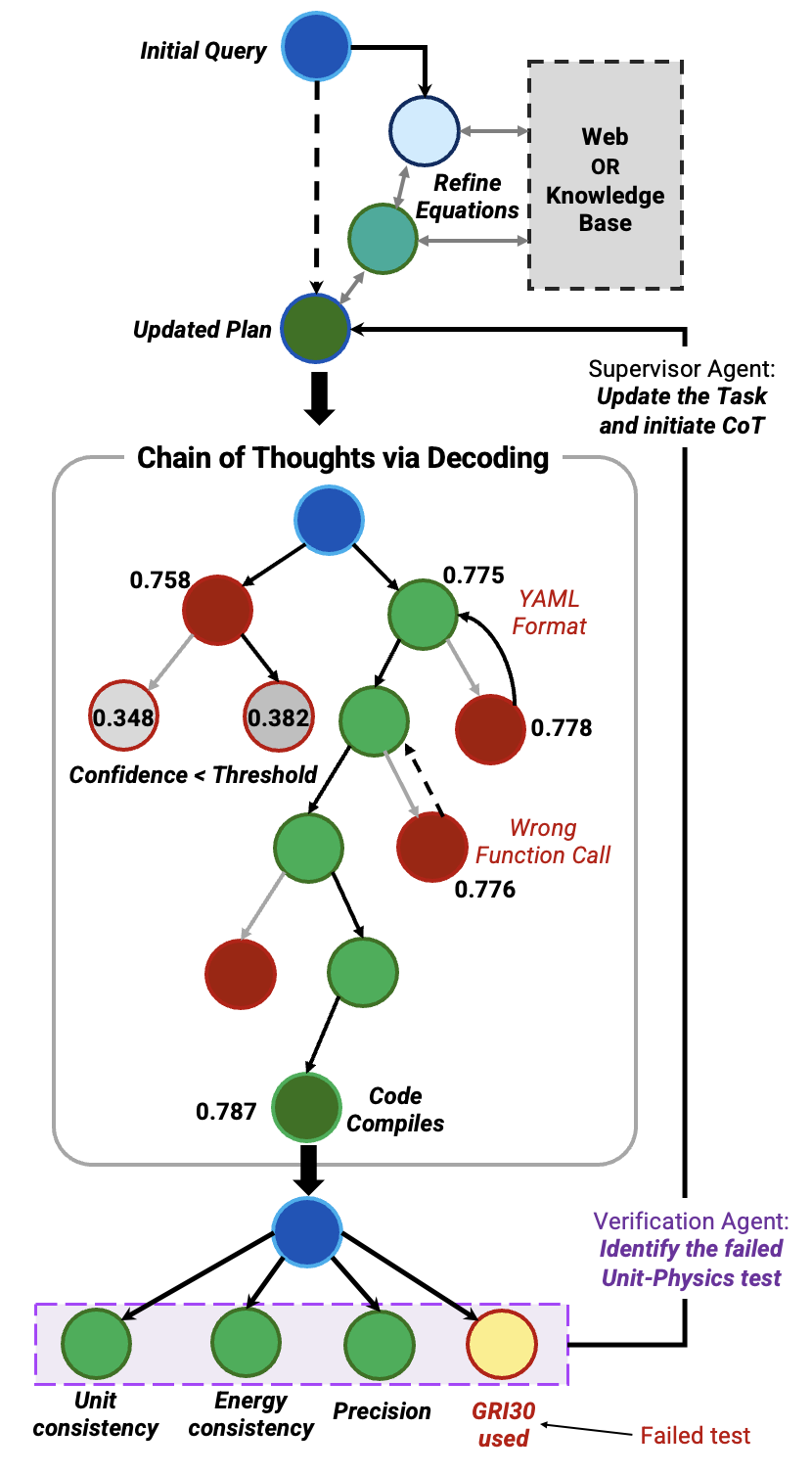}
\caption{Approximate states of the reactor task with primitives: (\tikzsymbol{fill=darkgreen}): correct state, (\tikzsymbol{fill=blue}): input to agent, (\tikzsymbol{fill=yellow}): mismatch detected, (\tikzsymbol{fill=gray}): pruned state and (\tikzsymbol{fill=red}): incorrect state. Numbers are CoT-confidence score self-reported by the Code agent.}
\label{fig:sol}
\end{figure}

The primitives-based AI system successfully synthesized the correct solution. Llama-3.3-70B model (sampling temperature of 0.6) supervised planning and tool use, while task-specific coding agents were instantiated with OSS-20B (sampling temperature of 0.3 and reasoning effort set to medium) for its code-generation proficiency. The system operated on a retrieval budget (at most two web queries), limiting dependence on external search and consistent with other evaluations in this study. Figure~\ref{fig:sol} summarizes the workflow as a state graph focusing on the solution and not agent blocks, where each node (colored circle) is a state or CoT step annotated with its confidence score (if available). Following the method in $\S$\ref{sec:method}, the initial query is passed to the supervisor agent that performs targeted fact/equation checks, formulates a plan, and communicates specific coding instructions to a code agent that explores four parallel candidates (one per GPU). The CoT states (in Fig.~\ref{fig:sol}) record progressive fixes across candidates—for example, correcting the Cantera mechanism format from \texttt{.cti} to \texttt{.yaml}. Candidates with confidence below 0.4 (user-defined) are pruned (gray states) and remaining candidates are passed to the 
Diagnostic agent. This agent performs an initial sanity check by executing the code and resolving issues relating to dependencies before proceeding, if needed.  
In our setup, the Python environment intentionally did not include Cantera. On execution, the agent correctly captured the traceback \texttt{ModuleNotFoundError: No module named `cantera'}. As a diagnostic repair, it issued a single command, \texttt{!pip install cantera}, to resolve the error \footnote{Execution privileges are explicitly granted for these runs; such commands should be used with caution.}.
Once the Diagnostic agent checks a candidate, the Verification Agent enforces the human-defined unit-physics tests. One primitive initially failed due to a mis-specified reaction mechanism (as shown). This failure is logged and stored in the graph database, and then fed back to the Supervisor Agent, which updates the plan and triggers another iteration. The system requires roughly 5-6 iterations to synthesize code that satisfies all checks. After these corrections, the primitive-grounded code runs successfully, yielding an IDT of 1.13179e-05 s. For statistical insights, this test case is repeated five times, obtaining four successful runs and one case where the system exhausts its iteration budget before converging.

\paragraph[Code Performance]{\textbf{Code Performance}\eatpunct}~\\
The code produced by the framework is evaluated against a reference implementation developed by a human expert and compared in three dimensions: (1) execution time, (2) memory usage, and (3) L$^{2}$ error. The representative chemical conditions with $\phi = 1$ and $p = 1$~atm with H$_{2}$–O$_{2}$ combustion and numerical conditions of $dt = 1e^{-10}$ with RK4 integrator are fixed and only the input temperature is varied from 1300 to 2400 K. 

\begin{figure}[t!]
\centering
\includegraphics[width=.35\textwidth]{ 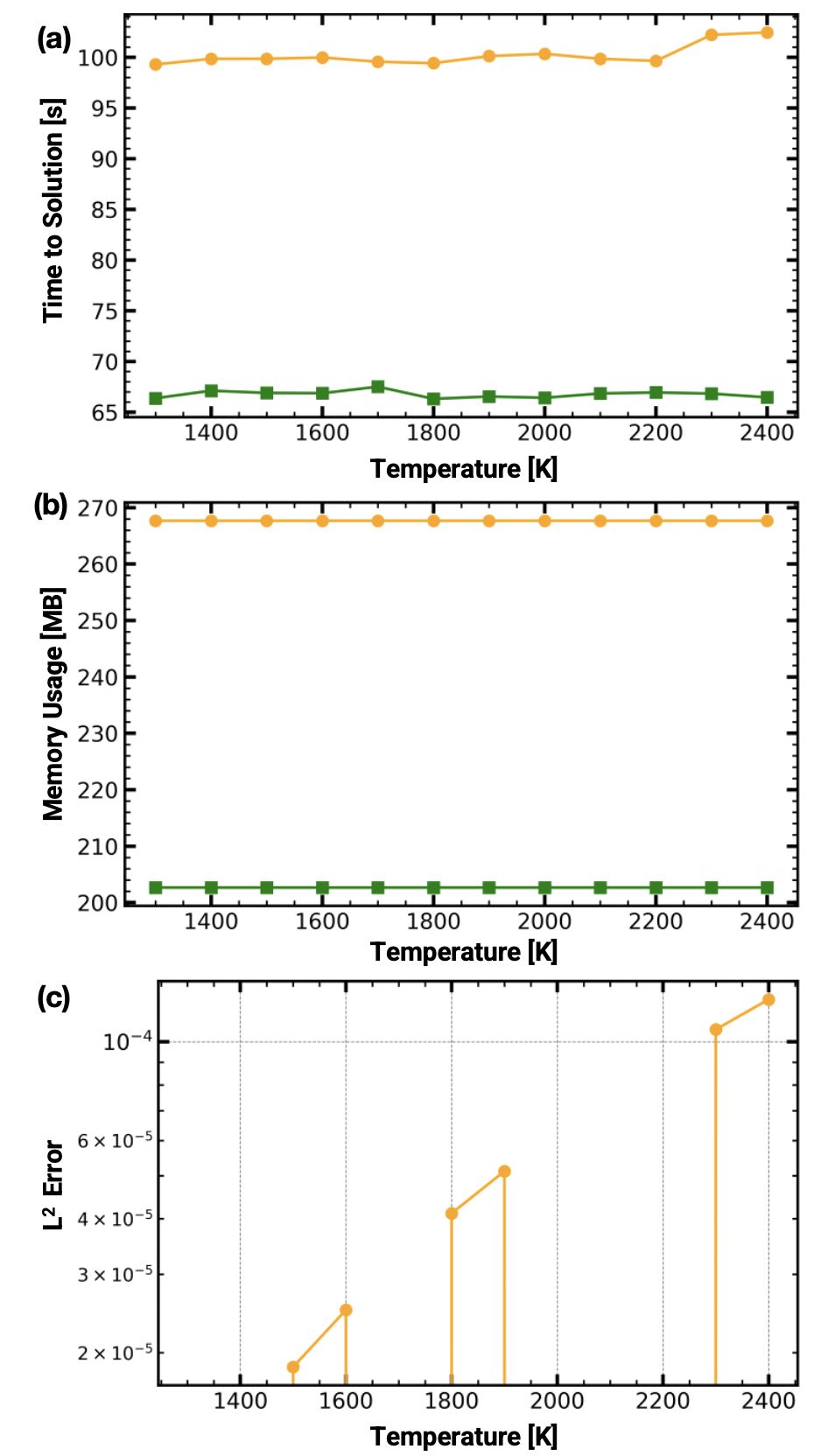}
\caption{Comparison between: (\tikzsymbol{fill=darkgreen}): Human+AI (Chain of Unit-Physics), and (\tikzsymbol{fill=orange}): Human, developed code implementation. Plots show (a) Time to solution,(b) Peak memory usage and (c) L$^2$ error of the Chain of Unit-Physics solution with respect to the human-expert reference (only single curve).}
\label{fig:coderuns}
\end{figure}

Figure~\ref{fig:coderuns} compares the performance of the proposed framework (green) with a reference code developed by a human expert (orange).
In terms of runtime [plot (a)], the human-expert code consistently achieves a longer time to solution—on the order of 32–34s (33.4\% on average) slower across the temperature range—indicating that the framework implementation is more optimized for wall-clock performance. The performance of the AI code can be attributed to using vectorized energy evaluations instead of explicitly looping over each species. Additionally, the proposed framework is more memory efficient [plot (b)], reducing peak memory usage from roughly 270 MB for the human code to about 200 MB (reduction of nearly 30\%), with only weak dependence on temperature. The accuracy of the generated solver is quantified by the L$^{2}$ error between the two solutions [plot (c)]. The error remains below $10^{-4}$ for all tested temperatures (mean relative error of $3.1\times10^{-3}$\%), with absolute match for some temperatures and a modest increase at higher temperatures, demonstrating that Chain of Unit-Physics closely reproduces the human-expert solution while trading a small increase in runtime for a substantial reduction in memory usage. Upon additional code review, the improved memory footprint of Chain of Unit-Physics arises from the way the AI-generated code organizes data: state variables are packed into a single contiguous structure rather than being split across multiple arrays and objects, which reduces overhead and allocator fragmentation. The slight slowdown in runtime was expected due to additional safeguard introduced by the model, an internal high temperature-bounds check (T $\geq$ 4000 K) that is frequently evaluated during the integration, however the vectorized approach offsets the time lost in checks. This extra validation step improves robustness, but adds a small computational penalty ($\sim$5s) relative to human-optimized implementation.

\paragraph[Cost Analysis]{\textbf{Cost Analysis}\eatpunct}~\\
The economic competitiveness and cost profile of the proposed framework are summarized in Table~\ref{tab:model_costs}. For the reference reactor task, Codex v0.44.0 consumes 352297 input and 16098 output tokens at a total API cost of \$0.25, while Claude Code v2.0.5 uses substantially fewer tokens (32110 / 2796) and therefore incurs a lower absolute cost of \$0.07. Normalizing by token count, however, reveals that Claude Code is actually more expensive on a per-token basis than Codex, so its apparent advantage is due primarily to greater token efficiency rather than cheaper pricing. In contrast, the proposed Chain-of-Unit-Physics framework processes 225502 input and 25382 output tokens on local GPUs, yielding a comparable or larger token budget than the commercial systems. The agent-level average token usage further indicates that most of this budget is concentrated in the Supervisor, Code, and Verification agents, with the Diagnostic agent contributing relatively less to the usage. These trends are consistent with the intended roles of the agents: the Supervisor oversees the entire process and therefore consumes the largest share of tokens, while the Code and Verification agents repeatedly iterate over the code once the Diagnostic agent has resolved dependency issues during the initial iterations.

\begin{table}[t!]
\centering
\begin{threeparttable}
\caption{Token usage and cost for different systems and agents.}
\label{tab:model_costs}

\small
% \footnotesize 
% \scriptsize
\begin{tabular}{lccc}\hline
%\textbf{Case}
\multicolumn{1}{c}{\begin{tabular}[c]{@{}c@{}}\textbf{AI}\\\textbf{System}\end{tabular}}
& \multicolumn{1}{c}{\begin{tabular}[c]{@{}c@{}}\textbf{Input}\\\textbf{Tokens}\end{tabular}}
& \multicolumn{1}{c}{\begin{tabular}[c]{@{}c@{}}\textbf{Output}\\\textbf{Tokens}\end{tabular}}
& \multicolumn{1}{c}{\begin{tabular}[c]{@{}c@{}}\textbf{Total}\\\textbf{Cost [\$]}\end{tabular}}
\\\hline

Codex v0.44.0         & 352\,297 & 16\,098 & 0.25 \\
Claude Code v2.0.5          & 32\,110  & 2\,796  & 0.07 \\
\textbf{Chain of Unit-Physics}\tnote{a} & 225\,502 & 25\,382 & - \\
\quad $\drsh$ Supervisor Agent  & 22\,050 & 1\,466 & - \\
\quad $\drsh$ Code Agent  & 8\,840 & 556 & - \\
\quad $\drsh$ Diagnostic Agent  & 1\,435 & 661 & - \\
\quad $\drsh$ Verification Agent  & 8\,365 & 2\,101 & - \\
\hline
\end{tabular}

% \footnotetext{Agent token count is averaged over successful runs.}
\begin{tablenotes}
\scriptsize
% \item[a] Cost from provider pricing as of Nov.\ 2025.
\item[a] Agent token count is averaged over successful runs.
\end{tablenotes}
\end{threeparttable}
\end{table}

\begin{figure}[t]
\centering
\includegraphics[width=.48\textwidth]{ 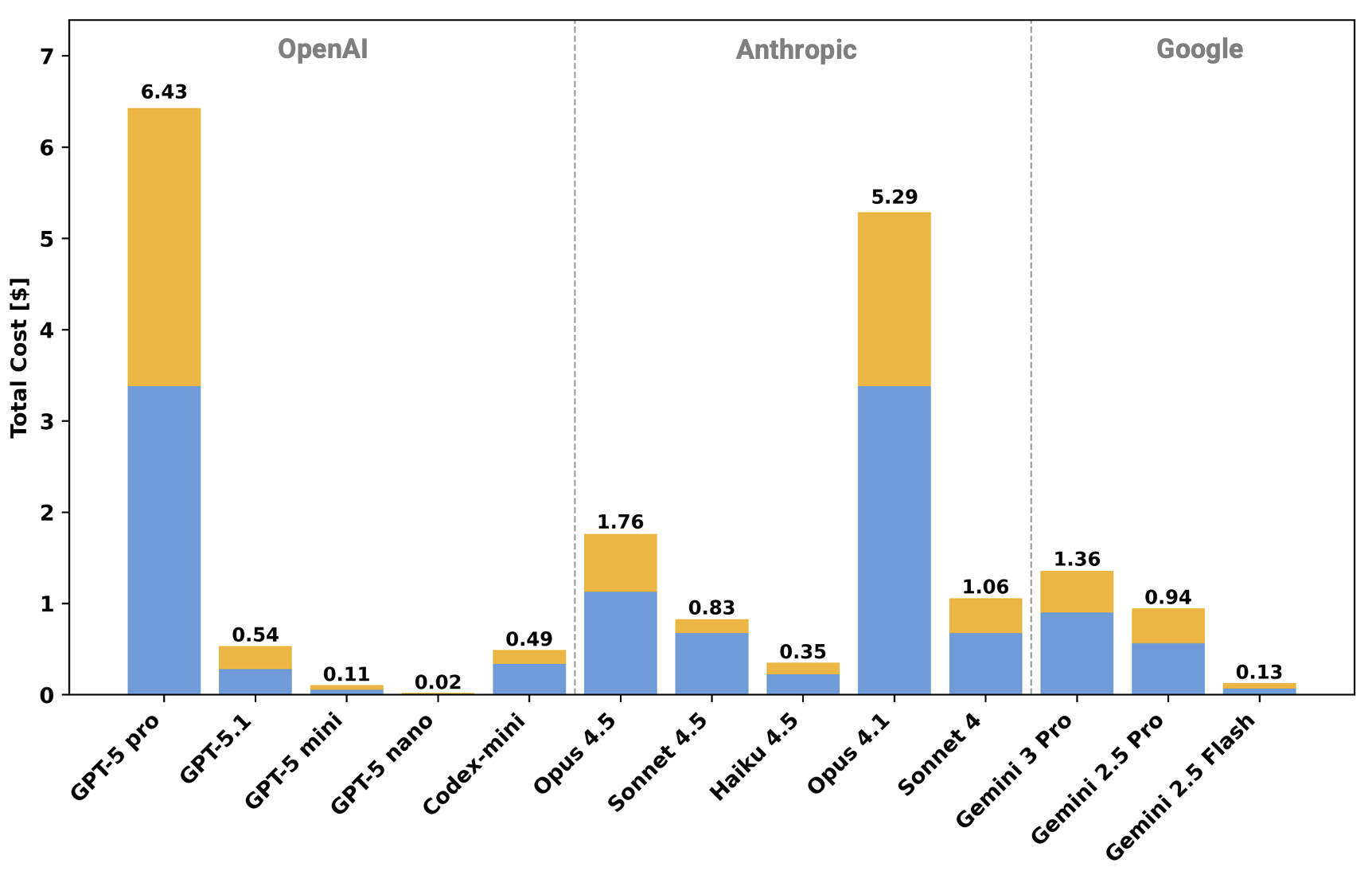}
\caption{Overall cost of the reactor task for different providers: (\tikzsymbol{fill=orange}): output token cost, (\tikzsymbol{fill=cyan}): input token cost. Numbers are based on pricing guides for respective providers as of Nov.\ 2025.}
\label{fig:cost}
\end{figure}

To further contextualize the API cost of the proposed framework relative to commercial providers, Fig.~\ref{fig:cost} breaks down the projected cost for different model variants. For the fixed Chain-of-Unit-Physics workload, the resulting API cost spans more than two orders of magnitude across providers. Among OpenAI and Google models, lightweight variants such as GPT-5 nano and Gemini 2.5 Flash are the most economical, at approximately \$0.02 and \$0.13 per run, respectively, while mid-sized models (GPT-5 mini, Codex-mini, Sonnet 4.5, Haiku 4.5, Gemini 2.5 Pro) fall in the \$0.1–\$1 range. In contrast, frontier chat models such as GPT-5 pro and Anthropic Opus are substantially more expensive, at \$6.43 and \$5.29 for the same token budget. Since Chain-of-Unit-Physics is implemented with mid-sized models, its effective cost is in the same regime as the mid-size hosted APIs, while remaining up to two to three orders of magnitude cheaper than the highest-end commercial offerings. Moreover, the Codex and Claude Code systems did not produce fully correct solutions for this task, so any practical deployment of those systems would likely require multiple retries or additional verification steps, further increasing their effective cost. Collectively, these observations suggest that the proposed framework is cost-competitive with commercial baselines for this workload, and that further gains in cost–effectiveness might come from techniques such as KV-cache compression~\cite{liu2024cachegen} or batch prompting~\cite{lin2024batchprompt} than from fundamental changes to the overall concept.

\section{Conclusion} 
This work examined ignition-delay computation for hydrogen combustion as a representative, high-stakes scientific task for agentic code generation. Closed-weight systems equipped with web access failed to produce a correct end-to-end solution. The experiments reveal four recurring error classes in agentic scientific coding: (i) interface hallucinations (nonexistent methods or attributes), (ii) overconfident assumptions about the scientific process (hard-wired logic that does not implement the correct algorithm), (iii) numerical and physical incoherence (e.g. invalid thermodynamic states), and (iv) configuration fragility (missing files or unsuitable default mechanisms). Across providers, models tended to select the same reaction mechanism, driven by its prominence in online documentation and frequent co-occurrence with ``Cantera" and ``ignition," rather than any explicit consideration of physical fidelity. Open-weight models display the same pattern: CoT decoding reduces interface hallucinations, but mainly shifts errors toward misuse of valid APIs.

The proposed Chain of Unit-Physics framework attains a correct solution in 5 to 6 iterations for the same ignition-delay problem; measured across five independent runs, four converged, and the remaining failure was attributable to an externally imposed token budget rather than to a modeling error. Performance analysis shows that the generated code closely matches the human-expert reference (L$^{2}$ error below 10$^{-4}$), with approximately 33.4\% faster runtime and using about 30\% less memory, a gain attributable to more compact data handling. When the same token budget is priced under different provider tariffs, the framework places in the cost band of mid-sized hosted models (on the order of \$0.1–\$1 per run), while avoiding excessive retries and verification passes that might be necessary for closed-model-based agents.

Overall, the results indicate that current agentic systems, even with tools and web access, are not yet reliable for this class of scientific workflow, and that embedding expert-designed unit-physics primitives as constraints is an effective way to improve reliability without sacrificing economic plausibility. The central idea is the use of portable physics and numerical primitives as an organizing scaffold for code search, rather than any specific underlying model family. This primitives-centric design offers a new horizon for human–AI collaboration in scientific computing: expert constraints define the admissible solution space, models search within that space, and resulting failures remain interpretable enough to train subsequent models. Future research should quantify how unit-test relaxation affects search and incorporate iterative refinement of the unit-physics tests \emph{in situ}, allowing the tests themselves to evolve within the code-generation cycle as new failure modes are discovered. Another direction for future work is to systematically evaluate influence of sampling temperature, an aspect not explored in the present study.

% A natural direction for future work is to study the \emph{strength} and \emph{granularity} of these constraints: if unit-physics tests are too weak, failure modes will slip through, but if they are too rigid, the model may focus on ``passing the tests" rather than solving the underlying scientific problem. 

\section*{Acknowledgments} 
This work was supported by the Center for Prediction, Reasoning, and Intelligence for Multiphysics Exploration (C-PRIME), a PSAAP-IV project funded by the Department of Energy, grant number DE-NA0004264 (program manager: Dr.\ David Etim).

%% Use \subsubsection, \paragraph, \subparagraph commands to 
%% start 3rd, 4th and 5th level sections.
%% Refer following link for more details.
%% https://en.wikibooks.org/wiki/LaTeX/Document_Structure#Sectioning_commands

% \appendix
% \section{Example Appendix Section}


\bibliographystyle{elsarticle-num} 
\bibliography{citations-refs}

\end{document}